\documentclass[prl,aps,showpacs,superscriptaddress,twocolumn]{revtex4}
\usepackage{graphicx}
\usepackage{amsmath}
\usepackage{amsfonts}
\usepackage{amssymb}
\usepackage{epsfig}
\begin{document}

\newcommand\lavg{\left\langle}
\newcommand\ravg{\right\rangle}
\newcommand\ket[1]{\left|#1\right\rangle}
\newcommand\bra[1]{\left\langle#1\right|}
\newcommand\braket[2]{\left.\left\langle#1\right|#2\right\rangle}
\def\I {{\rm 1} \hspace{-1.1mm} {\rm I} \hspace{0.5mm}}

\title{
Local control of entanglement in a spin chain}

\author{Francesco Plastina}\affiliation{ Dip. Fisica, Universit\`a della
Calabria, \& INFN - Gruppo collegato di Cosenza, 87036 Arcavacata
di Rende (CS) Italy}  \author{Tony J. G. Apollaro}
\affiliation{Dip. Fisica, Universit\`a degli Studi di Firenze, Via
G. Sansone 1, 50019 Sesto F.no (FI) Italy}
\date{\today}

\begin{abstract}
In a ferromagnetic spin chain, the control of the local effective
magnetic field allows to manipulate the static and dynamical
properties of entanglement. In particular, the propagation of
quantum correlations can be driven to a great extent so as to
achieve an entanglement transfer on demand toward a selected site.
\end{abstract}
\bigskip
 \maketitle When the
superposition principle is applied to an (at least) bi-partite
system, highly non local and purely quantum correlations
(entanglement) can appear among the parties. This constitutes a
crucial resource for many applications in quantum
communication\cite{bdv}. Consequently, the problem of entanglement
distribution has become of central interest: quantum correlations
are generated by local interactions; therefore methods are
required to transfer either the entangled particles or their state
at a distance. It has been shown theoretically that spin chains
are efficient quantum channels for short distance entanglement
transfer \cite{bose}; thus the ability to manipulate the
propagation of entanglement in a spin chain can be very important
and it has already been shown that breaking the translational
invariance of the chain can produce very interesting results in
this respect \cite{chris,alt}.

In this paper, we show that a control of both static and dynamical
properties of entanglement can be achieved by acting locally to
modify the level spacing of some qubits. In the magnetic language,
we analyze a system subject to a spatially inhomogeneous magnetic
field; that is, a system with `diagonal defects' (`impurities'),
\cite{lfsantos}.

In contrast to the usual case in which local actions cannot affect
non-local physical quantities; here, due to the spin-spin
interaction, the local control of the magnetic field modifies both
the distribution of correlations in the ground state and the
entanglement propagation along the chain. Indeed, spatial
inhomogeneities of the external field lead to an Anderson-like
localization of entanglement \cite{anderson}, and this can occur
even for a single defect\cite{economou}, giving rise to a
mirror-like effect in the entanglement propagation \cite{noi}.
Contrary to the usual description of localization phenomena, we do
not conceive the impurity just as a bit of disorder in the system,
but rather intend the modification of the local level spacing as a
knob to {\it i}) control the content of ``static'' (ground state)
entanglement, and {\it ii}) drive its propagation along the chain.

We consider a 1-D XX spin-$\frac{1}{2}$ closed chain, placed in an
external magnetic field which is homogeneous everywhere but for
two defect sites $l_1$ and $l_2$. This model is described by the
Hamiltonian $H=H_0 + H_{def}$, with
\begin{eqnarray}
&& H_0=
-\frac{\omega_0}{2}\sum_{i=-\frac{N}{2}}^{\frac{N}{2}}{\sigma}_{z}^{i}
-J\sum_{i=-\frac{N}{2}}^{\frac{N}{2}}
({\sigma}_{x}^{i}{\sigma}_{x}^{i+1}
+{\sigma}_{y}^{i}{\sigma}_{y}^{i+1}) \\ && H_{def} =
-\frac{1}{2}\sum_{j=1,2} \alpha_j {\sigma}_{z}^{l_j}
\label{hamiltonian}\end{eqnarray} where $\omega_0$ is the level
spacing of each qubit, except for those residing at sites $l_i$,
which have level separation $\omega_0+ \alpha_i$. Henceforth, we
shall take $J=1$ and use the ferromagnetic coupling constant as
our energy unit (we have also set $\hbar=1$). Furthermore, we set
to zero the energy of the completely separable eigenstate $\ket
0^{\otimes N}$, which is the unperturbed ground state for
$\omega_0 >1$, and which is still an energy eigenstate in the
presence of defects.

This model can be solved exactly via the Jordan-Wigner (JW)
transformation, which maps the spin chain into a spin-less fermion
system \cite{apol}. However, we do not need the general solution
here, since we will deal with states having at most one tilted
spin (in the JW language, states lying in the single particle
subspace), \cite{altri}.  In the continuum limit $N\rightarrow
\infty$, we solve the model (restricted to the single particle
sector of its Hilbert space) using the Green operator technique
\cite{economou}.

To this end, we consider the Green operator $G_0$ describing the
homogeneous chain. It is known \cite{economou,noi} that $G_0$
displays a branch cut on the real axis in the complex energy plane
for energies $E \in [\omega_0-1,\omega_0+1]$. This cut signals a
continuous energy band, which survives even in the presence of the
defects.

Given $G_0$, the full (single particle) Green operator is
\begin{equation}
G=G_0 + G_0 T G_0,\end{equation} where the T-matrix is given by
\begin{equation}
T= \frac{\sum_i \ket{l_i} t_i \bra{l_i} + \ket{l_2} t_1
G_0(l_2,l_1) t_2 \bra{l_1} + \mbox{h.c.}}{1- t_1 t_2 G_0(l_1,l_2)
G_0(l_2,l_1)}, \label{tmatr}
\end{equation}
where $\ket n = \ket{0}^{\otimes N-1} \otimes \ket{1}_n$ is the
state with one spin down (or, equivalently, one fermionic
excitation) located at site $n$, and where the scattering
coefficients are
$$t_i = -\frac{\alpha_i}{2+ \alpha_i G_0(l_i,l_i)}.$$
The T-matrix describes multiple scattering events of the single
particle excitation at the two defects, and it is precisely the
re-summation of the various scattering amplitudes which gives rise
to the denominator in Eq. (\ref{tmatr}). The existence of zeros of
this denominator is crucial since it implies that, besides the
energy band discussed above, the model with defects displays some
(at most two, in fact) discrete energy levels.

They lay above or below the energy band depending on the values of
the defect fields, and their eigen-energies can be obtained
analytically in some special cases. For example, for nearest
neighbors defects, with distance $d=|l_2-l_1|=1$, by setting
$x_{loc}=E_{loc}-\omega_0$ one gets
$$
x_{loc}= \frac{\alpha_1 \alpha_2 \frac{\alpha_1+\alpha_2}{4} \pm
\sqrt{\left [1+\frac{(\alpha_1-\alpha_2)^2}{4}\right ](1-
\frac{\alpha_1 \alpha_2}{2})^2}}{1-\alpha_1 \alpha_2}
$$Only one of these two solutions (the one with the
plus sign in front of the square root) exists if the defect
strengths satisfy the relation $\alpha_1^{-1}+\alpha_2^{-1}\geq
1$. That is, the parameter space is divided in two regions,
characterized by the number of discrete energy eigenvalues (which
can be one or two).

If we restrict ourselves to the ordered phase of the unperturbed
chain, $\omega_0 >1$, and if we consider only $\alpha_i>0$, the
lowest (or the only existing) localized level becomes the
fundamental state of the perturbed problem.

Many of these features obtained for nearest neighbors defects are
generic and do not depend on their distance. From the analytic
properties of the Green operator, \cite{economou}, one can show
that i) at most two localized states are present, whose position
with respect to the energy band depends on the sign of the
$\alpha$'s, ii) a region in the $\alpha_1$-$\alpha_2$ plane exists
in which only one discrete state is found. However, this
one-eigenstate region becomes thinner and thinner as $d$
increases, iii) the localized states always have the structure of
single excitation states of the form $\ket{\psi_{loc}}= \sum_n b_n
\ket{n}$. The amplitudes $b_n$, obtained from the residues of $G$,
can be expressed in terms of an inverse localization length
$\xi=-\ln[-x_{loc} -\sqrt{x_{loc}^2-1}]$, as a function
``bi-localized'' around the two defects:
\begin{equation}
b_n = \mbox{const} \Bigl (K_1 e^{-\xi |n-l_1|} + K_2 e^{-\xi
|n-l_2|} \Bigr ). \label{wavefu} \end{equation} The coefficients
$K_i$ are given by $$K_i= \left ( \frac{\alpha_i}{2
\sqrt{x_{loc}^2-1}-\alpha_i} \right )^{\frac{1}{2}} \, .$$ Their
ratio indicates the relative weights of the two localization
region. For $\alpha_1\gg \alpha_2$ one gets $|K_2|\ll |K_1|$ and
the discrete levels are localized around $l_1$, while for
$\alpha_1\ll \alpha_2$ the localization center is $l_2$ since
$|K_2|\gg |K_1|$. For equal defect strengths, $\alpha_1=\alpha_2$,
one finds $K_2 = \pm K_1$, where the upper (lower) sign refers to
the lower (higher) of the two eigenstates. Thus, if
$\alpha_1=\alpha_2$, the two discrete states are given by
equal-weight coherent superpositions of two localized parts
centered on the two impurities. These states are highly entangled
and display strong quantum correlations between the defects and
their neighborhoods.

Even for quite small values of the defect fields, the localization
length is smaller than the inter-site spacing (already
$\alpha_1=\alpha_2 =1$ gives $\xi^{-1}<1$, for any distance
$d=|l_1-l_2|$). This implies that the two discrete eigen-states
are approximately given by the Bell combinations $(\ket{01}\pm
\ket{10})$, with the rest of the chain almost completely
factorized in the state $\ket 0$.
\begin{figure}
\includegraphics[width=0.4\textwidth]{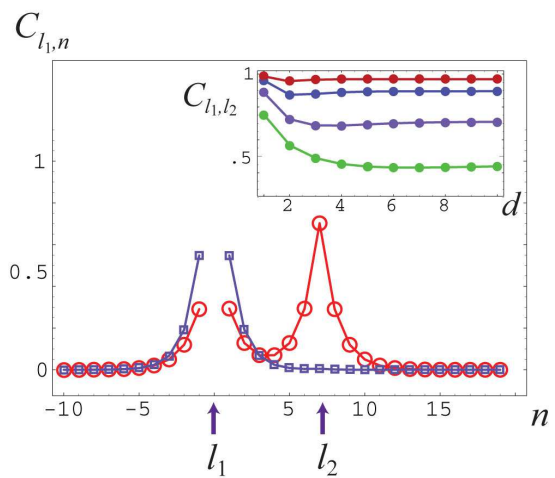}
\caption{(Color online) Ground state concurrence between the
$l_1$-th and the $n$-th spin of the chain. The entanglement can be
remotely controlled by changing the local field at the other
defect. The line with open circles (boxes) corresponds to
$\alpha_1=\alpha_2=1.5$ ($\alpha_1=2, \alpha_2=1.5$). For equal
defect strengths, spin $l_1$ is entangled both with its own
neighborhood and with the other defect's one. In the asymmetric
case, quantum correlations only survive within the localization
region. The inset shows the concurrence between the two defects as
a function of their distance for various values of defect fields.
From below to above, the plots correspond to $\alpha_1=\alpha_2=
0.25 ; 0.5; 1; 2.$ }\label{stat}
\end{figure}
It is noteworthy that this structure does not depend on the
distance between the two defects, see the inset of Fig.
\ref{stat}, where the concurrence $C_{l_1l_2} = 2 |b_1 b_2|$
between the defects is shown as a function of $d=|l_1-l_2|$. Thus,
a long distance bi-partite entanglement can be obtained in the
ground state of the chain. A similar behavior has been found in
Ref. \cite{campos}, with the difference that in our case this is a
bulk property rather than a surface effect.

For generic values of the defect amplitudes, these states display
entanglement between any pair of spins residing near each of the
two defects, with the peculiarity that the pairwise entanglement
for two spins around the same defect depends on the value of the
local magnetic field at the other defect. This is illustrated in
figure (\ref{stat}), where it is shown that the entanglement
around $l_1$ is modified by changing the strength of local field
at $l_2$, thus achieving  a remote entanglement control.

The bi-local character of the discrete levels strongly affects the
transport of entanglement along the chain. In particular, we
consider the possibility of using the chain to send one partner of
a maximally entangled pair. We assume that the spin at the sender
site $s$ is prepared in a singlet state with an external
(un-coupled) qubit. The interaction between the spins causes a
transfer of entanglement along the chain. Ideally, after a given
transmission time $t$, one would like to get a singlet between the
external qubit and the one residing at a receiving site $r$. To
characterize the quality of the transmission, we evaluate the
(final) concurrence between the external and the $r$-th qubits,
denoted by $C_{r}(t)$.

This is given by $C_r(t)=|f_{s\rightarrow r}(t)|$, \cite{bose},
where $f_{s\rightarrow r}(t)$ is the amplitude for the transfer of
a fermionic excitation from site $s$ to $r$. This is expressed in
terms of the retarded Green operator as
\begin{equation}
f_{s\rightarrow r} = \sum_{E_{loc}} e^{-i E_{loc} t}  b_r b_s^* +
\int_{-\pi}^{\pi} \!\! d \theta e^{-i E t} g_r(E) g_s^*(E).
\label{effers}\end{equation} The first term describes transport
mediated by localized states, while the second one gives a
spin-wave mediated transfer, \cite{njp}. We analyze them
separately.

\begin{figure}
\includegraphics[width=0.45\textwidth]{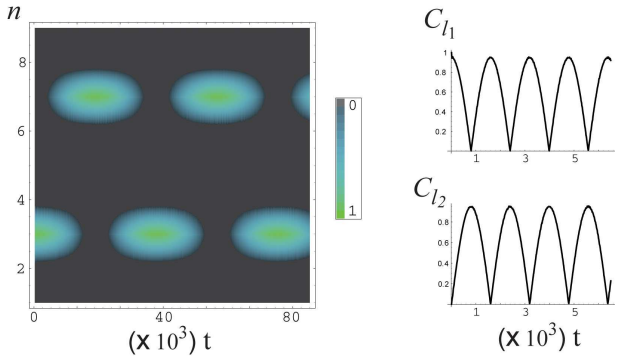}\\
\caption{(color online) Rabi oscillations of the entanglement
between the two defects for the case $\alpha_1= \alpha_2=1.5$.
}\label{rabiosc}
\end{figure}

\begin{figure}
\includegraphics[width=0.5\textwidth]{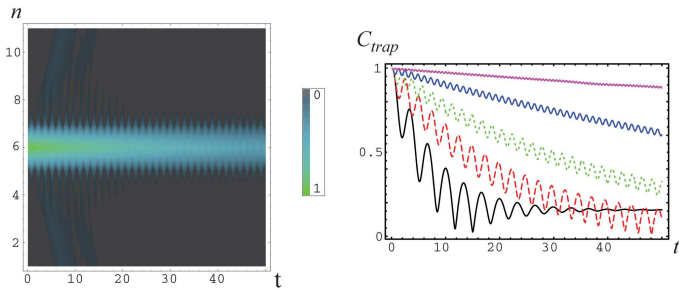}\\
\caption{(color online) Left: Entanglement trapping between the
two defects in the case $\alpha_1= \alpha_2=2$. Right: Time
dependence of the trapped concurrence for $\alpha_1= \alpha_2 =
0.5,1,1.5,2.5,5$ (from below to above). }\label{trappi}
\end{figure}

The first contribution is effective only within a region of length
$\xi^{-1}$ around the two defect sites. This gives the noteworthy
possibility of transmitting from one neighborhood to the other.
This effects is illustrated in Fig. (\ref{rabiosc}), where the
entanglement is shown to jump from one defect to the other.
Indeed, if the sender site coincides with one of the defects (say
$l_1$), than the system evolves by performing almost perfect Rabi
oscillations between the two discrete localized states, with a
Rabi frequency given by $\omega_R=(E_{loc 1}-E_{loc 2})$. This can
be understood by noticing that the initial singlet state between
$l_1$ and the external qubit can be approximately written as
$$\ket{\psi_{in}}  \simeq \frac{1}{\sqrt 2} \Bigl [ \ket{0_{l_1}}
\ket{1_{ext}} -\frac{1}{\sqrt 2} ( \ket{\psi_{loc 1}} +
\ket{\psi_{loc 2}} ) \ket{0_{ext}} \Bigr ], $$ which implies that
the concurrence between the $l_i$-th qubit and the external one is
an harmonic function:
\begin{equation}
C_{l_1}(t) \simeq |\cos \omega_R t|, \qquad C_{l_2}(t) \simeq
|\sin \omega_R t|.
\end{equation}
The second term in Eq. (\ref{effers}) is an integral over the
energy band, parameterized as $E=\omega_0- \cos \theta$. It
contains the state amplitudes of the continuous energy band, which
can be written in terms of the retarded Green and $T$ operators:
$$g_n(E) = \bra n \Bigl [ \I + G_0^{+} T^+ \Bigr ] \ket{\psi_0
(E)},$$ with the un-perturbed states such that $\lavg n|\psi_0(E)
\ravg = e^{i n \theta}/\sqrt{2 \pi}$. These states represent
distorted spin waves of the system. They are the stationary
scattering states of the single-particle Hamiltonian and can be
constructed, starting from the usual magnon excitations, by
including corrections due to multiple scattering at the defects.
\begin{figure}
\includegraphics[width=0.35\textwidth]{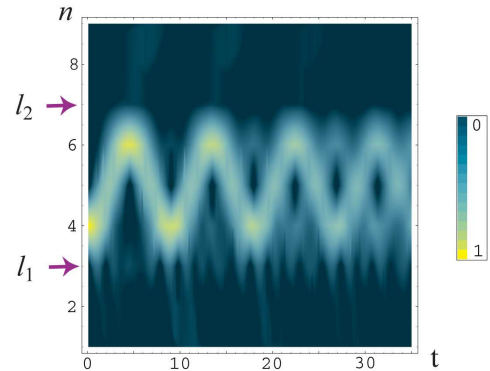}\\
\caption{(color online) Entanglement bouncing between the two
defects which act like (non-perfect) mirrors. The location of the
impurities is indicated by the two arrows and the defect-strengths
are $\alpha_1= \alpha_2=1.5$. }\label{bouncev}
\end{figure}
For moderate values of the $\alpha$'s, the distortion is not that
big and the unperturbed plane wave nature can still be recognized.
As a result, the energy eigenstates form (approximate) standing
waves between the two defects. Their pattern is reflected in the
entanglement propagation shown in Fig. (\ref{bouncev}) where the
sender site is located between the two impurities which act as
potential barriers for the spin waves, so that entanglement
bounces back and forth between these two mirrors. The extreme
situations is reached when the defects are next to nearest
neighbors to each other, thus realizing an entanglement trap, see
Fig. (\ref{trappi}). Since the mirrors are not perfect,
\cite{noi}, the trapped entanglement decreases with time, the
superimposed time oscillations being due to the $e^{-i E_{loc}t}$
factor of the symmetric discrete eigenstate (the only one that
matters, in this case). Once these oscillations are subtracted,
the short time behavior of the concurrence is found to be
parabolic, with a convexity that decreases as the defect
amplitudes are increased. A long-time (residual) trapped
entanglement is also present, which is due to the tails of the
localized state. It diminishes with increasing $\alpha$'s as the
localization length does, and becomes negligible if $\xi^{-1}$ is
much smaller than the site spacing.

The Rabi oscillations illustrated above give a mean to transfer
the entanglement reversibly between the defects. This method works
independently of their distance, but has the drawback that the
Rabi period increases with distance as $T \sim \alpha^d$ (this can
be derived by perturbation theory for large $\alpha$'s, see Ref.
\cite{lea}).

The form of the discrete states, however, suggests another, more
effective method to achieve entanglement transfer on demand
between the two defects; namely, the adiabatic passage (see
\cite{adiaa} for a related proposal in which the coupling strength
is varied instead of the local field).

The idea is to change the defect strengths adiabatically, so that
the ground state of the system, having the general form given in
Eq. (\ref{wavefu}), is adiabatically changed from the initial
state $\ket{\psi_{gs}(i)} \simeq \ket{l_1}$ localized at $l_1$, to
the final state $\ket{\psi_{gs} (f)} \simeq \ket{l_2}$, localized
at $l_2$. This can be done by modifying the defect fields from the
initial values
$$\alpha_1(i) \gg 1, \quad \alpha_2(i) \ll 1 \qquad \Rightarrow
\quad b_n(i) \simeq \delta_{n,l_1},$$ to the final (reversed) ones
$$\alpha_1(f) \ll 1, \quad \alpha_2(f) \gg 1 \qquad \Rightarrow
\quad b_n(f) \simeq \delta_{n,l_2}.$$ If this is done
adiabatically, the system always remains in its instantaneous
ground state, thus realizing the desired entanglement transfer
provided the initial singlet state involves the external and the
$l_1$-th qubit. To ensure the adiabaticity, the rates of change of
the $\alpha$'s have to bo much smaller than the difference between
the energies of the two discrete levels, $E_{21} = E_{loc
2}-E_{loc 1}$. The most dangerous point in this respect (i.e., the
smallest $E_{21}$) occurs at the crossing, when
$\alpha_1(t)=\alpha_2(t)$. But, since at this point $E_{21} \sim
\alpha^d$, if the adiabatic procedure is designed such that the
crossing occurs for a very small $\alpha$, than the adiabaticity
can be preserved even for transfer times much smaller than the
Rabi period. This procedure has the additional advantage of
effectively decoupling the receiving site from the rest of the
chain after the transfer has been performed due to its large final
local field.

To summarize, we have discussed how to manipulate a spin chain
with local control fields, showing that the static entanglement
can be remotely controlled, and that entanglement propagation can
be adjusted to a large extent in order to achieve transfer on
demand. One possibility to implement this model in a realistic set
up is to use the method proposed in Ref. \cite{duan} to "engineer"
spin chains with atoms in an optical lattice. The addition of
external static local electric or magnetic fields should enable
the control of the qubit energy level spacing which is essential
to test our proposal.

This work has been supported by MIUR through the PRIN2005029421
project.


\begin{thebibliography}{99}
\bibitem{altri}
Since we work in the single particle subspace, properties of
entanglement very similar to those derived here are obtained with
the XXZ model. For the sake of clarity, we limited ourselves to
the simplest case (XX spin chain).
\end{thebibliography}
\end{document}